\documentclass[useAMS,usenatbib]{mn2e}
\usepackage{graphicx}
\usepackage{xcolor}
\usepackage{amssymb,amsmath,bm}
\usepackage{hyperref}
\usepackage[export]{adjustbox}
\hypersetup{colorlinks=true,citecolor=blue}
\usepackage{multirow}

\hypersetup{
    colorlinks=false,
    citecolor=blue,
    linkcolor=blue,
    linktoc=page
}

\newcommand{\aap}{Astron.\ Astrophys.\ }
\newcommand{\apj}{Astrophys.\ J.\  }
\newcommand{\apjl}{Astrophys.\ J.\ Lett.\  }
\newcommand{\apjs}{Astrophys.\ J.\ Suppl.\  }
\newcommand{\mnras}{Mon.\ Not.\ R.\ Astron.\ Soc.\ }
 
\newcommand{\nat}{Nature} 
 
\newcommand{\physrep}{{\em Phys.\ Rep.}}
\newcommand{\pasj}{Publ.\ Astron.\ Soc.\ Japan\ }
\newcommand{\prd}{Phys.\ Rev.\ D\ }
\newcommand{\jcap}{Journal of Cosmology and Astro-Particle Physics}

\newcommand{\avg}[1]{\ensuremath{\left\langle \,#1\, \right\rangle}}

\newcommand{\tj}[6]{ \begin{pmatrix}
   #1 & #2 & #3 \\
   #4 & #5 & #6 
  \end{pmatrix}}

\newcommand{\be}{\begin{equation}}
\newcommand{\ee}{\end{equation}}

\newcommand{\bea}{\begin{eqnarray}}
\newcommand{\eea}{\end{eqnarray}}
\newcommand{\bdm}{\begin{displaymath}}
\newcommand{\edm}{\end{displaymath}}

\renewcommand{\vec}{\mathbf}

\def\Mpc{\, h^{-1} \, {\rm Mpc}}

\newcommand{\eq}[1]{Eq.~(\ref{#1})}
\newcommand{\eqs}[2]{Eqs.~(\ref{#1},\ref{#2})}
\newcommand{\fig}[1]{Figure~\ref{#1}}

\newcommand{\vb}[1]{\mathbf{#1}}
\def\ie{{\em i.e.}~}
\def\eg{{{\em e.g.}~}}

\title[Wide angle effects in the Zeldovich approximation]{The Zeldovich approximation and wide-angle redshift-space distortions}

\author[E. Castorina, M. White]
{Emanuele Castorina$^{1,2}$\thanks{e-mail: ecastorina@berkeley.edu}, 
Martin White$^{1,2}$\thanks{e-mail: mwhite@berkeley.edu}
 \\~\\
\footnotesize
\footnotesize
$^1$Berkeley Center for Cosmological Physics, University of California, Berkeley, CA 94720, USA\\
$^2$Lawrence Berkeley National Laboratory, 1 Cyclotron Road, Berkeley, CA 93720, USA\\
}

\begin{document}
\maketitle 

\begin{abstract}
The contribution of line-of-sight peculiar velocities to the observed redshift of objects breaks the translational symmetry of the underlying theory, modifying the predicted 2-point functions.  These `wide angle effects' have mostly been studied using linear perturbation theory in the context of the multipoles of the correlation function and power spectrum .
In this work we present the first calculation of wide angle terms in the Zeldovich approximation, which is known to be more accurate than linear theory on scales probed by the next generation of galaxy surveys.
We present the exact result for dark matter and perturbatively biased tracers as well as the small angle expansion of the configuration- and Fourier-space two-point functions and the connection to the multi-frequency angular power spectrum.  We compare different definitions of the line-of-sight direction and discuss how to translate between them.  We show that wide angle terms can reach tens of percent of the total signal in a measurement at low redshift in some approximations, and that a generic feature of wide angle effects is to slightly shift the Baryon Acoustic Oscillation scale.
\end{abstract}

\section{Introduction}
\label{sec:intro}

The clustering of galaxies observed in redshift surveys exhibits anisotropies due to the contribution of line-of-sight velocities to the measured redshift \citep{Kai87,H92,H98,Pea99,Dod03}.  This effect, known as redshift space distortions (RSD), enables us to constrain the rate of growth of large-scale structure \citep{Guzzo08,PerWhi08}, test theories of modified gravity \citep{Joyce15} and probe the constituents of the Universe such as massive neutrinos \citep{Lesgourgues,Wei13}.

When modeling the 2-point statistics of redshift-space clustering, most analyses make the ``plane parallel approximation'', where the direction of the line-of-sight of each object in a pair is assumed to be the same.  This means the redshifts of the two objects receive contributions from the same component of the velocity, which increases the symmetry of the system and allows simple Fourier and configuration space analyses.  This approximation is usually excellent on small scales and for deep surveys.  However for surveys which are relatively shallow and probe large scales -- or interferometers whose primary beams cover large sky areas -- there are ``wide angle'' effects which need to be considered.  There is a large literature examining these effects within linear perturbation theory  \citep{H92,HC96,H98,ZH96,Sza98,Sza04,Dat07,PapSza08,ShaLew08,Bonvin,Raccanelli,YS15,Slepian15,Rei16,CasWhi18} but relatively little analytic work beyond linear theory.  In this paper we present an analysis of wide angle effects to lowest order in Lagrangian perturbation theory, i.e.~the Zeldovich approximation (ZA; \citealt{Zel70}).  The ZA is quite accurate on large scales, and provides a better modeling of features in the spectrum such as baryon acoustic oscillations than does linear theory \citep{Tas14a}.  It also contains more angular structure than is present in linear theory, allowing us to examine in more detail the impact of relaxing the plane parallel approximation.
A calculation of non-linear wide angle terms in standard (Eulerian) perturtabion theory beyond linear theory was presented in \citet{ShaLew08}. Our analysis differs from theirs in that we resum the linear displacement fields and include the effect of biasing of halos and galaxies with respect to the dark matter field.

In this work we shall focus on the `physical' wide angle terms, arising from the projection of peculiar velocities onto a line of sight that varies across the sky, and their effect on the non linear dynamics of the matter and galaxy density fields. It is well known that a survey's selection function will introduce extra wide angle terms \citep{Kai87,HamCul96,Sza98,CasWhi18}. The latter have been studied in linear perturbation theory, but we are not aware of any formalism to include them beyond this regime. However if the selection function is slowly varying with redshift we expect higher order terms to be small.

The outline of the paper is as follows.  In \S\ref{sec:background} we review the formalism of redshift-space distortions and the wide angle effects and outline the geometry of the problem, including the bisector and end-point conventions for defining the line of sight.  In this section we also provide an explicit connection between the 3D correlation function and power spectrum multipoles and the multi-frequency angular power spectrum (MAPS), showing explicitly that the angular structure arises through projection and allowing computation of the MAPS beyond the plane-parallel approximation.
In \S\ref{sec:zeldovich} we describe how to include wide angle effects into the ZA for both the correlation function \citep{Tas14b} and the power spectrum. The full result is evaluated numerically in \S\ref{sec:za}, whereas in \S\ref{sec:ZAx2} we present a small angle expansion and the relation between the ZA, linear theory and the different conventions for the line of sight.
We conclude in \S\ref{sec:conc} and discuss some technical details in the appendices.

\section{Wide angle effects}
\label{sec:background}

The study of ``wide angle'' effects in galaxy clustering has a long history.  Here we briefly review some important background material and make connections between different conventions for the line of sight and different 2-point clustering statistics.  We shall adopt the notation and conventions of \citet{CasWhi18}, which builds upon the earlier work of \citet{HC96,ZH96,Sza98} and \citet{PapSza08}.

\begin{figure}
\begin{center}
\begin{picture}(250,250)
\Large
\thicklines
\qbezier(125.0,0.0)(125.0,112.5)(125.0,218.8)
\qbezier(125.0,0.0)(156.2,125.0)(187.5,250.0)
\qbezier(125.0,0.0)(93.8,93.8)(62.5,187.5)
\qbezier(62.5,187.5)(125.0,218.8)(187.5,250.0)
\thinlines
\multiput(125,219)(0,6){5}{\line(0,1){3}}
\put(110,60){$\displaystyle{\frac{\theta}{2}}$}
\put(128,60){$\displaystyle{\frac{\theta}{2}}$}
\put( 70,120){$\vec{s}_1$}
\put(160,120){$\vec{s}_2$}
\put(130,180){$\vec{d}$}
\put( 70,182){\rotatebox{27}{$s(1-t)$}}
\put(160,227){\rotatebox{27}{$st$}}
\put(115,205){$\phi$}
\put(160,250){\vector(-2,-1){75}}
\put(110,230){$\vec{s}$}
\end{picture}
\caption{The assumed geometry and angles.  The two galaxies lie at $\vec{s}_1$
and $\vec{s}_2$, with separation vector $\vec{s}=\vec{s}_1-\vec{s}_2$ and
enclosed angle $\theta$.
We take the line of sight to be parallel to the angle bisector, $\vec{d}$,
which divides $\vec{s}$ into parts of lengths $st$ and $s(1-t)$.  The
separation vector, $\vec{s}$, makes an angle $\phi$ with the line of sight
direction, $\hat{d}$. \label{fig:conf}}
\end{center}
\label{fig:triangle}
\end{figure}
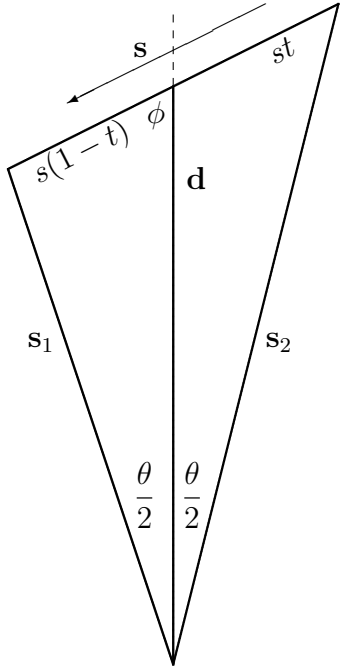
\subsection{Geometry}

The geometry of a generic redshift space configuration for the 2-point function is shown in Fig.~\ref{fig:triangle}.    Three numbers are enough to properly describe the system, and several possible choice of parametrization are available.
The observer $O$ is looking at two objects at $\vb{s}_1$ and $\vb{s}_2$, separated by $\vb{s} = \vb{s}_1 - \vb{s}_2$ and the line of sight (LOS) $\vb{d}$ is defined as the bisector of the angle $\theta$ between $\vb{s}_1$ and $\vb{s}_2$.  The LOS and the separation vector form an angle $\phi$, with $\cos(\phi)\equiv \mu$. 
We also define $t\in [0,1]$ through
\begin{equation}
\vb{s}_1 = \vb{d} + (1-t)\, \vb{s} \quad , \quad
\vb{s}_2 = \vb{d} - t \,\vb{s} \quad .
\end{equation}
The clustering is independent of the direction of $\vec{d}$ and of rotations of $\vec{s}$ about $\vec{d}$, thus the correlation function can be written as a function of $s$, $\mu$ and $d$.  We follow \citet{Rei16} and define a small parameter $x\equiv s/d\ll 1$, as this will prove convenient later.

\citet{CasWhi18} show that for the configuration in Fig.~\ref{fig:triangle}
\begin{equation}
  \vec{d} = \frac{s_1s_2}{s_1+s_2}\left(\hat{s}_1+\hat{s}_2\right)
  \quad , \quad
  d^2 = \frac{4s_1^2s_2^2}{(s_1+s_2)^2}\cos^2\frac{\theta}{2}
\end{equation}
and
\begin{equation}
  1-t = \frac{-1+\mu x+\sqrt{1+\mu^2 x^2}}{2\mu x}
  \simeq \frac{1}{2} + \frac{\mu x}{4} - \frac{\mu^3 x^3}{16} + \cdots
\end{equation}
Another popular choice of coordinate system, which we call the end-point parametrization, is to use one of the two galaxies, for instance $\vec{s}_1$, as the line of sight. This is mostly motivated by the fact that in this case estimators for the power spectrum multipoles can be written as simple Fast Fourier Transforms (FFT) \citep{Sco15,Bia15,Han17}. On the other hand one loses the reflection symmetry around the observer, \ie the redshift space correlation function is not invariant under $\mathbf{s} _1 \rightarrow -\mathbf{s} _1$. For this choice of LOS wide angle effects are much bigger than for the midpoint or the bisector, and odd multipoles are also generated \citep{Rei16,CasWhi18}. It is relatively straightforward to go from one parametrization to another one using
\begin{align}
  \hat{s}_1\cdot\hat{s}_2 &\simeq 1 - \frac{x_1^2}{2}(1-\mu_1^2) + \cdots \\
  \mu &\simeq \mu_1 - \frac{x_1}{2}(1-\mu_1^2) + \cdots \\
  x   &\simeq x_1\left(1 + \frac{\mu_1x_1}{2}\right) + \cdots
\end{align}
with $\mu_1=\hat{s}\cdot\hat{s}_1$ and expansion parameter $x_1=s/s_1\ll 1$.  We shall develop the formalism within the bisector convention and discuss how to transform to the end-point convention where appropriate.

\subsection{Correlation function}

It is common to express the correlation function as an expansion in $\mu$ via:
\begin{eqnarray}
\label{eq:xiell}
\xi_s(s,d,\mu) &=& \sum_{\ell} \xi_\ell(s,d) \mathcal{L}_{\ell}(\mu)  \\
 &=& \sum_{\ell n} x^n\xi_\ell^{(n)}(s)\mathcal{L}_{\ell}(\mu)
\end{eqnarray} 
where $\mathcal{L}_\ell$ indicates the Legendre polynomial of order $\ell$ and in the second line we have expanded $\xi_\ell(s,d)$ in powers of the wide angle parameter, $x=s/d$.  The lowest order terms are the familiar ``plane parallel'' approximation.  In linear theory \citep{H92}
\begin{align}
\label{eq:xiK}
\xi_{0}^{pp}(s) &= \xi_0^{(0)}(s) \left(1+\frac{2}{3}f + \frac{1}{5}f^2 \right)\notag \\
\xi_{2}^{pp}(s) &= \xi_2^{(0)}(s)\left(-\frac{4}{3}f - \frac{4}{7}f^2\right) \notag \\
\xi_{4}^{pp}(s) &= \xi_4^{(0)}(s)\left(\frac{8}{35}f^2 \right)
\end{align}
with
\begin{equation}
  \xi_\ell^{(0)}(s) = \int\frac{k^2\,dk}{2\pi^2}\ P(k)j_\ell(ks)
\label{eqn:xi_ell_n}                      
\end{equation}
and $P(k)$ the linear theory power spectrum.

\subsection{Power spectrum}

The definition of the redshift space power spectrum requires some care if we drop the plane-parallel approximation \citep{ZH96}.
What is always well-defined is the ``local'', \ie LOS-dependent, power spectrum \citep{Sco15,Rei16},
\begin{equation}
  P(\vb{k},\vb{d}) \equiv \int \mathrm{d}^3 s
    \, \xi(\vb{s},\vb{d})e^{-i\vb{k}\cdot\vb{s}}
\end{equation}
which can be expanded in multipoles as
\begin{align}
  P(\vb{k},\vb{d}) &= \sum_{\ell} P_\ell(k,d)
  \mathcal{L}_\ell\left(\hat{k}\cdot\hat{d}\right)
\label{eqn:Pkd_expansion} \\
 &\equiv \sum_n (kd)^{-n} P_\ell^{(n)}(k)\mathcal{L}_\ell\left(\hat{k}\cdot\hat{d}\right)
 \quad .
\label{eqn:Pln_bisector}
\end{align}
In this case multipoles of the correlation function and of the power spectrum are still related by a Hankel-transform,
\begin{align}
\label{eq:Pkell}
  P_\ell(k,d) &= 4\pi(-i)^\ell\int\,s^2\,\mathrm{d}s
  \ j_\ell(ks)\,\xi_\ell(s,d)
\end{align}
and thus
\begin{equation}
  P_\ell^{(n)}(k) = 4\pi(-i)^\ell\int s^2\,\mathrm{d}s\ (ks)^{n}\,\xi_\ell^{(n)}(s)\,j_\ell(ks)
\label{eqn:Pln_bisector_hankel}
\end{equation}
with inverse
\begin{equation}
  \xi_\ell^{(n)}(s) = \int\frac{k^2\,\mathrm{d}k}{2\pi^2}\ (ks)^{-n}P_\ell^{(n)}(k)\,j_\ell(ks)
  \quad .
\end{equation}

In observations the most commonly used estimator for the power spectrum multipoles has been proposed by \cite{Yam06}:
\begin{align}
\label{eq:PkY}
  \hat{P}_L^Y(k) &\equiv \frac{(2L+1)}{V}\int \frac{\mathrm{d}\Omega_\vb{k}}{4\pi}\mathrm{d}^3 s_1 \mathrm{d}^3 s_2 \notag \\
  & \times \delta(\vb{s}_1) \delta(\vb{s}_2) e^{-i\vb{k}\cdot \vb{s}} \mathcal{L}_L\left(\hat{k}\cdot\hat{d}\right)
\end{align}
where $V$ is the survey volume, and the line of sight can be either the bisector or the midpoint.
Taking the expectation value of this estimator yields \citep{CasWhi18}
\begin{align}
\avg{\hat{P}_L^Y(k)}
  &=(2L+1) \int\frac{\mathrm{d}^3d}{V}\,\frac{\mathrm{d}\Omega_\vb{k}}{4\pi}\,P(\vb{k},\vb{d}) \mathcal{L}_L(\hat{k}\cdot\hat{d}) \\
  &= \int\frac{\mathrm{d}^3d}{V} \,P_L(k,d) \quad .
\label{eqn:PY_Pkd}
\end{align}
At lowest order in $x$ the expectation value of $\hat{P}_L^Y$ is simply $P_\ell^{(0)}(k)$ as in Eq.~(\ref{eqn:Pln_bisector_hankel}).

For computational reasons the estimator is more commonly defined using the direction to one of the two galaxies as the line of sight
\begin{align}
\label{eq:PkFFT}
  \hat{P}_L^{FFT}(k) &\equiv \frac{(2L+1)}{V}\int \frac{\mathrm{d}\Omega_\vb{k}}{4\pi}\mathrm{d}^3 s_1 \mathrm{d}^3 s_2 \notag \\
  & \times \delta(\vb{s}_1) \delta(\vb{s}_2) e^{-i\vb{k}\cdot \vb{s}} \mathcal{L}_L\left(\hat{k}\cdot\hat{s}_1\right)
\end{align}
which can be evaluated using FFTs \citep{Sco15,Bia15,Han17}, compared to brute force pair sum required in \eq{eq:PkY}. The ensemble average of the FFT estimator can be related to the multipoles of the theoretical correlation function defined using the bisector or the end-point as the line of sight following \cite{CasWhi18}, who have also shown that wide angle corrections to \eq{eq:PkFFT} are in general much bigger than the one of \eq{eq:PkY}.  We shall denote the coefficients of the expansion of the end-point-based $P_\ell(k,s_1)$ in powers of $(ks_1)$ as $\tilde{P}_\ell^{(n)}(k)$ to distinguish them from the bisector-based $P_\ell^{(n)}(k)$ of Eq.~(\ref{eqn:Pln_bisector}). 

\subsection{Fourier-Bessel expansion and MAPS}

A third representation of the 2-point function is in terms of the Fourier-Bessel (sFB) expansion \citep{Lahav94,Fisher94,HT95,Percival04,Pad01,Pratten13,CasWhi18} or its configuration space analog, the multi-frequency angular power spectrum (MAPS; \citealt{Dat07}; see also \citealt{CasWhi18}).  The latter has most commonly been used to describe fluctuations measured by wide-area, interferometric, $21\,$cm instruments \citep[e.g.][]{Sha14}.

In this formalism, one describes the triangle of Fig.~\ref{fig:triangle} in terms of the two side lengths ($s_1$ and $s_2$) and the enclosed angle ($\theta)$.  Expanding the $\theta$-dependence in Legendre polynomials
\begin{equation}
\label{eq:Cls}
  \xi(\vec{s}_1,\vec{s}_2) = \sum_{\ell=0}^\infty  \frac{2\ell+1}{4\pi} C_\ell(s_1,s_2)\mathcal{L}_\ell(\cos\theta)
\end{equation}
the coefficients, $C_\ell$, are the MAPS and their one dimensional Hankel transform along $s_1$ and $s_2$ the angular power spectra
\begin{align}
\label{eq:Clk}
  C_\ell(k_1,k_2) = \int s_1^2\mathrm{d}s_1\, \int s_2^2\mathrm{d}s_2\, C_\ell(s_1,s_2) j_\ell(k_1 s_1) j_\ell(k_2 s_2)\;.
\end{align}
So far most treatments of spherical power spectra have been confined to linear theory, limiting their use in real data where one has to deal with \eg Fingers-of-God (FOG) \citep{Percival04} and non-linearities in the galaxy field \citep{Pratten13}. This was due to the very complicated expression the spherical coefficients of the density fields take beyond linear theory.

However, we know from the 3D Cartesian analysis that at large scales most of the power is confined in a few multipoles, $L$, and thus the complex structure of the MAPS or angular power spectra must predominantly result from projection effects. This is similar to what happens with the CMB where only a few multipoles are relevant at recombination and the rich structure we observe today is due to LOS projection \citep{Dod03}.  Let us try to see this directly from \eqs{eq:Cls}{eq:Clk}.  Inverting \eq{eq:Cls}
\begin{equation}
  C_\ell(s_1,s_2) = 2\pi \int_{-1}^{1} \mathrm{d}(\cos\theta)\ \xi(\vec{s}_1,\vec{s}_2)\mathcal{L}_\ell(\cos\theta)
\end{equation}
we see that if we approximate $\mathcal{L}_\ell(\cos\theta)$ with $J_0(\ell\theta)$, valid for $\theta\ll 1$, then the $C_\ell(s_1,s_2)$ is the LOS Fourier transform of $P(\vec{k})$ with $\ell=k_\perp d$ (Appendix \ref{app:flatsky_MAPS}).

In general the MAPS can be expressed as an integral over the power spectrum.  The full expression\footnote{One proceeds by using the addition theorem and Rayleigh expansion of the plane wave, then expands the $Y_{\ell m}$ using solid harmonics as in Appendix E of \citet{CasWhi18}, and simplifies the angular integrals using $3j$-coefficients.} using the bisector definition of $\vec{d}$ is extremely cumbersome, but it simplifies dramatically if we instead use the end-point definition.  Expanding the exponentials using the Rayleigh expansion of the plane wave and combining products of spherical harmonics using the Gaunt integral one can show
\begin{align}
 & 2\pi\int\mathrm{d}(\cos\theta)\ \mathcal{L}_\ell(\cos\theta)
   \int\frac{\mathrm{d}\Omega_k}{4\pi} e^{i\vec{k}\cdot(\vec{s}_1-\vec{s}_2)}
   \mathcal{L}_L(\hat{k}\cdot\hat{s}_1) \nonumber \\
 &= 4\pi\sum_{\lambda} (2\lambda+1)  i^{\lambda-\ell}
 \tj{\lambda}{\ell}{L}{0}{0}{0}^2j_{\lambda}(ks_1)j_{\ell}(ks_2) \ .
\end{align}
Thus our $C_\ell$ become
\begin{align}
\label{eq:ClsT}
  C_\ell(s_1,s_2) &= 2\pi \int\mathrm{d}(\cos\theta)\mathcal{L}_\ell(\cos\theta)\int\frac{\mathrm{d}^3k}{(2\pi)^3}\ e^{i\vec{k}\cdot(\vec{s}_1-\vec{s}_2)} \nonumber \\
  &\times \sum_{L}P_L(k,s_1)\mathcal{L}_L(\hat{k}\cdot\hat{s}_1) \\
  &= \sum_{L\lambda} F^{\ell}_{L\lambda}\int\frac{k^2\mathrm{d}k}{2\pi^2}\, P_{L}(k,s_1) j_{\lambda}(ks_1) j_\ell(ks_2)
\end{align}
where
\begin{align}
  F^\ell_{L\lambda} &= 4\pi(2\lambda+1)i^{\lambda-\ell}\tj{\ell}{L}{\lambda}{0}{0}{0}^2
\end{align}
The triangle condition of the $3j$-coefficients makes the sum over $\lambda$ finite, as $|\lambda-\ell|\le L$.

Note that we did not have to assume any specific model for the power spectrum: \eq{eq:ClsT} can be used to describe any model of the angular correlation function.  The final expression is remarkably simple and shows how the structure at $\ell > L$ is only due to projection along the direction to the two galaxies. 
The expression for the angular power spectrum can be written in an even simpler form,
\begin{equation}
\label{eq:ClkT}
  C_\ell(k_1,k_2) = \sum_{L\lambda} \frac{F^{\ell}_{L\lambda}}{4\pi} \int s^2\mathrm{d}s\,P_{L}(k_2,s) 
   j_\ell(k_1s)j_{\lambda}(k_2s)
\end{equation}
and expanding $P_L(k_2,s)$ in powers of $(k_2s)$ \citep{Rei16,CasWhi18}
\begin{align}
\label{eq:ClkT2}
  C_\ell(k_1,k_2) &= \sum_{L\lambda n} \frac{F^{\ell}_{L\lambda}}{4\pi} \tilde{P}_L^{(n)}(k_2)
  \int s^2\mathrm{d}s\,(k_2s)^{-n} j_\ell(k_1s)j_{\lambda}(k_2s) \notag \\
  &\equiv \sum_{L\lambda n} \mathcal{M}^{(n)}_{\ell,L,\lambda}(k_1,k_2) \tilde{P}_L^{(n)}(k_2)
\end{align}
Hence the sFB power spectrum is the product of the (3D) multipoles of the power spectrum times a geometric term that can be expressed\footnote{The relevant expression is on p.~401 (\S13.4) of \citet{Watson}.} in terms of hypergeometric functions and does not depend on any cosmological parameters.  Note for small $L$ and large $\ell$ the $F^\ell_{L\lambda}$ are non-zero only for $\lambda\approx\ell$ and the integral is highly peaked around $k_1=k_2$\footnote{For $\ell\ne\lambda$ and $n=0$ the off-diagonal terms in \eq{eq:ClkT2} decay as $\text{min}[(k_1/k_2)^\ell,(k_2/k_1)^\lambda]$ when $\ell$ and $\lambda$ are large.}.
The above expressions provide an exhaustive description of the two point statistics for redshift surveys in spherical coordinates, completing the description of \cite{Liu16} for Intensity Mapping surveys and of \cite{Pas17} for imaging surveys, neither of which included redshift space distortions.
The other main issue in spherical analysis is the estimate of the covariance matrix. Since $\ell$ can easily go up to a few hundreds and the power spectrum is estimated in tens of $k$-bins, the dimensionality of the covariance makes the problem very quickly intractable \citep{Percival04}.
\eqs{eq:ClsT}{eq:ClkT2} offer a simple solution to this problem, as the $C_\ell$ and the $P_L$ are linearly related to each other by a matrix that can be `inverted' to find an optimal data compression. Given a survey geometry and galaxy selection function one needs to measure only the $\ell$, $k_1$, $k_2$ that maximize the signal and at the same time keep the dimensionality of the problem low enough.
We finally point out that \eq{eq:ClkT2} provides an elegant and unbiased way to remove systematics in the plane of the sky, \eg fiber collision in spectroscopic instruments \citep{Hahn17}, that by definition affect only the low-$k_\parallel$ modes. The wave numbers appearing on the left-hand side are indeed radial Fourier modes, whereas the ones of the right-hand side are 3D Cartesian modes.

\section{Zeldovich approximation}
\label{sec:zeldovich}

Almost all prior work on wide angle effects used an Eulerian, linear theory description of the 2-point function.  Instead we shall base our analytic model of wide angle effects on $1^{\rm st}$ order Lagrangian perturbation theory -- the ZA \citep{Zel70}.  Despite the more than 40 years since it was first introduced, the ZA still provides one of our most accurate models for the distribution of cosmological objects.  It has been applied to understanding the impact of non-linearities on BAO \citep{PWC09,Noh09,McCSza12,TasZal12a}, to reconstruction \citep{TasZal12b,Whi15}, as the basis of an effective field theory \citep{Por14,VWA15} and as a rapid means of simulating large-scale structure \citep{Dor80,Col93,PauMel95,SahCol95,Hid14,Chuang15}.  The ZA can easily incorporate wide angle effects, and is quite accurate on the large scales where such effects are most important.
For a pedagogical introduction to the ZA and the analytic calculation of the correlation function see e.g.~\citet{CLPT,Whi14,Tas14b}.

\begin{figure}
\begin{center}
\resizebox{0.95\columnwidth}{!}{\includegraphics{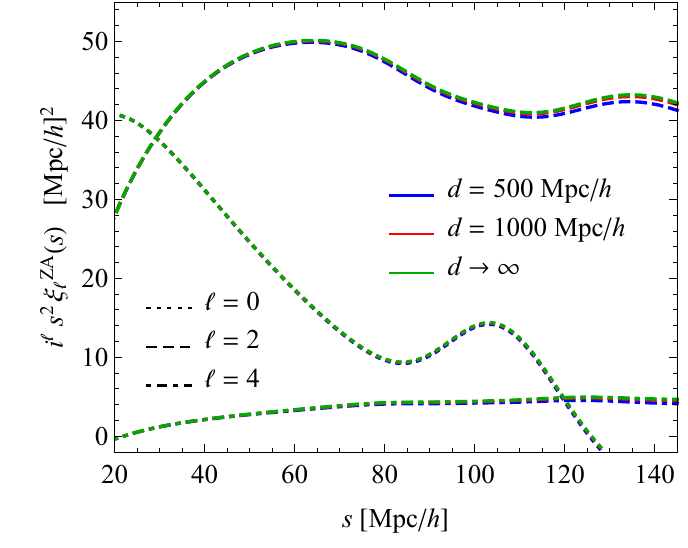}}
\end{center}
\caption{The multipoles of the redshift-space, matter correlation function ($b=1$), computed from Eq.~(\ref{eqn:ZelXi}), with the bisector definition of the line-of-sight, for $d=500\,h^{-1}$Mpc (blue), $1\,h^{-1}$Gpc (red) and $10\,h^{-1}$Gpc (green).  The differences are barely visible on this linear scale and the $10\,h^{-1}$Gpc lines form an excellent approximation to the plane-parallel or small-angle limit ($d\to\infty$).}
\label{fig:multipoles_ZA}
\end{figure}

\begin{figure}
\begin{center} 
\resizebox{.9\columnwidth}{!}{\includegraphics{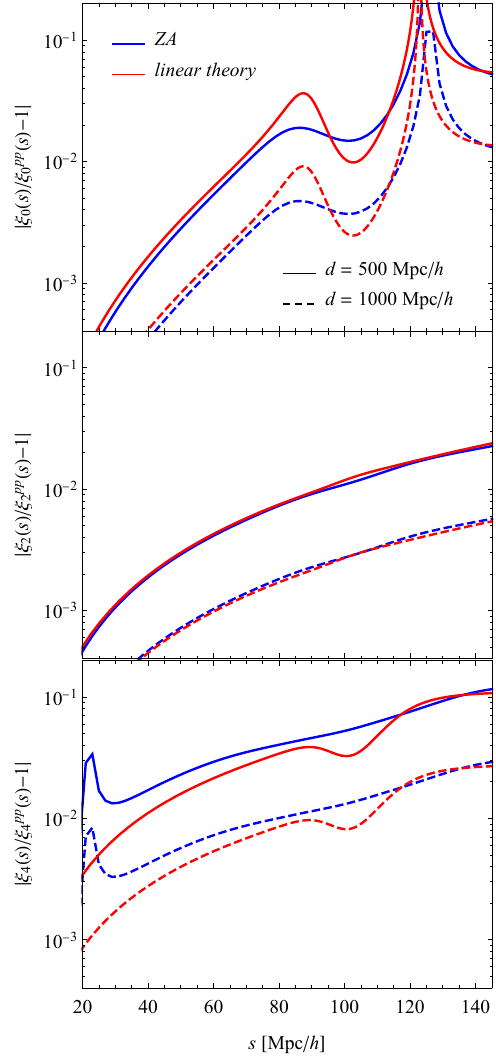}}
\end{center}
\caption{The ratio of the multipoles of the redshift-space, matter correlation function ($b_1=0$ or $b=1$) for $d=500\,h^{-1}$Mpc (solid) and $1\,h^{-1}$Gpc (dashed) to the same multipoles in the plane parallel approximation (Eq.~\ref{eq:xiK}).  The blue lines show the ZA (Eq.~\ref{eqn:ZelXi}) while the red lines show linear theory \citep[from][]{CasWhi18}.}
\label{fig:multipoles_b1}
\end{figure}

\begin{figure}
\begin{center}
\resizebox{.9\columnwidth}{!}{\includegraphics{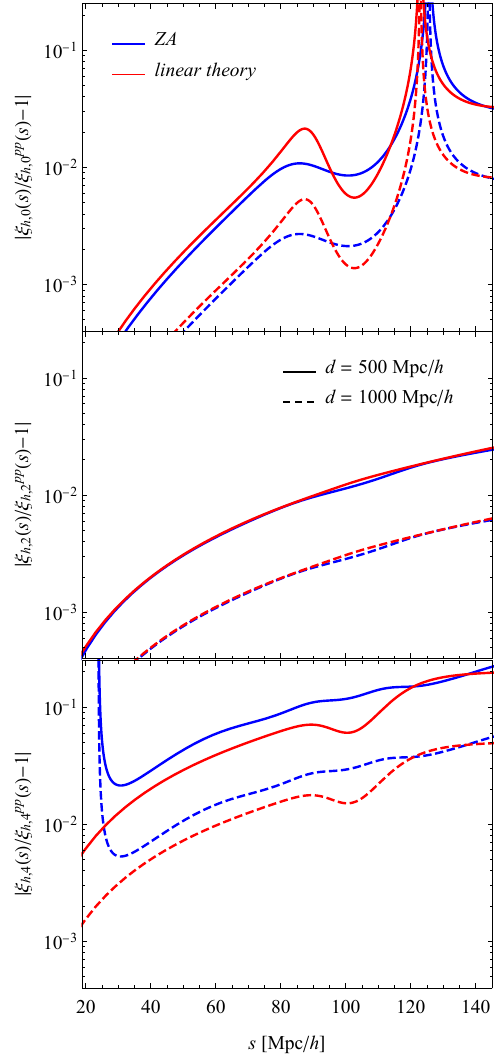}}
\end{center}
\caption{Same as Fig.~\ref{fig:multipoles_b1} but now for tracers with linear bias $b=2$ (i.e.~$b_1=1$).}
\label{fig:multipoles_b2}
\end{figure}

Following the usual approach we denote the Lagrangian position of a fluid element by $\vec{q}$, and its final (Eulerian) position by $\vec{x}=\vec{q}+\vec{\Psi}(\vec{q},t)$.  This serves to define the displacement, $\vec{\Psi}$.  To lowest order in perturbation theory, i.e.~the ZA,
\begin{equation}
  \vec{\Psi}(\vec{q}) = \int\frac{d^3k}{(2\pi)^3} e^{i\vec{k}\cdot\vec{q}}
  \ \frac{i\vec{k}}{k^2}\delta_L(\vec{k})
  \quad .
\end{equation}
Within the ZA the transition to redshift space is straightforward, and indeed this is one of the major advantages of Lagrangian perturbation theory in large-scale structure.  For an object observed in direction $\hat{s}$ the redshift-space displacement is related to the real-space displacement by
\begin{equation}
  \vec{\Psi}(\vec{q})\to \mathbf{R}\vec{\Psi}(\vec{q})
  = \left(1+f\hat{s}\hat{s}\right)\vec{\Psi}(\vec{q})
\end{equation}
For our purposes the important point to note is that $\hat{s}$ is $q$-independent.  This means that it is fixed in the integrals over $d^3q$ that define the density field and correlation function (see below) which makes the inclusion of wide-angle effects within the ZA very straightforward \citep{Tas14b}.

\begin{figure}
\resizebox{.9\columnwidth}{!}{\includegraphics{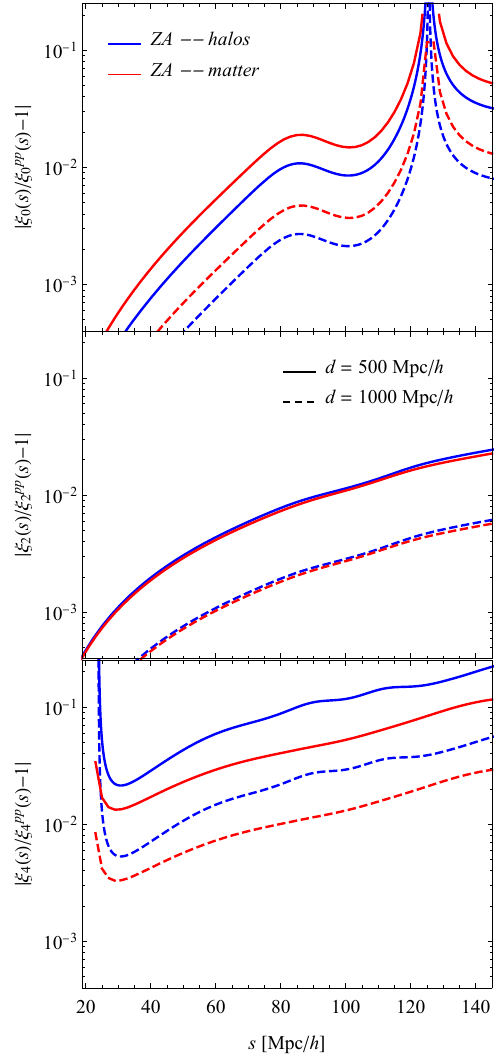}}
\caption{Comparison between the size of the wide angle terms (in the ZA) for dark matter (red) and tracers with $b=2$ (blue).}
\label{fig:multipoles_bZA}
\end{figure}

If we make the standard definitions that $\vec{\Delta}=\vec{\Psi}(\vec{q}_1)-\vec{\Psi}(\vec{q}_2)$ where $\vec{q}=\vec{q}_1-\vec{q}_2$ and $A_{ij}=\langle\Delta_i\Delta_j\rangle=X(q)\delta_{ij}+Y(q)\hat{q}_i\hat{q}_j$, $U_i=\langle\Delta_i\delta_2\rangle=U(q)\hat{q}_i$, $\xi=\langle\delta_1\delta_2\rangle$ then
\begin{eqnarray}
  X(q) &=& \int\frac{dk}{2\pi^2} P(k) \left[\frac{2}{3}-2\frac{j_1(kq)}{kq} \right] \\
  Y(q) &=& \int\frac{dk}{2\pi^2} P(k)\left[-2j_0(kq)+6\frac{j_1(kq)}{kq} \right] \\
  U(q) &=& \int\frac{dk}{2\pi^2} P(k)\left[-k\,j_1(kq)\right] \\
  \xi(q)&=&\int\frac{dk}{2\pi^2} P(k)\left[k^2j_0(kq)\right]
\end{eqnarray}
where $P(k)$ is the linear theory power spectrum, and the redshift-space correlation function can be written as
\begin{align}
  1 + \xi^{(s)}(s,d,\mu) &= \int d^3q\frac{d^3k}{(2\pi)^3}
  \ e^{ik_i(q-s)_i}\ e^{-(1/2)k_iA_{s,ij}k_j} \nonumber \\
  &\times \left[ 1 + 2ib_1k_iU_{s,i}+b_1^2\xi + \mathcal{O}(P_L^2)\right] \nonumber \\
  &= \int\frac{d^3q}{(2\pi)^{3/2}|A_s|^{1/2}}
  \ e^{-(1/2)(\vec{q}-\vec{s})A_s^{-1}(\vec{q}-\vec{s})} \nonumber \\
  &\times \left[ 1 - 2 b_1g_iU_i + b_1^2\xi + \mathcal{O}(P_L^2)\right]
\label{eqn:ZelXi}
\end{align}
where
\begin{eqnarray}
  A_s &=&  (R_1^2 + R_2^2 ) \Sigma^2 \delta_{ij} +\left( R_1 \xi R_2 + R_2\xi R_1 \right) \\ 
  A_s &=& \frac{1}{2}\left( R_1AR_2 + R_2AR_1 \right) \\
  g_i &=& A_{s,ij}^{-1}\left(q_j-s_j\right) \\
  U_{s,i} &=& \frac{1}{2}\left(R_{1,ij}+R_{2,ij}\right)U_j
\end{eqnarray}
In the above we have written the real space  displacement correlation function as $A_{ij} = 2 \Sigma^2 \delta_{ij} + 2 \xi_{ij}$ and included biased tracers to lowest order in the Lagrangian bias, following \citet{Mat08}.  This will be sufficient for our purposes (see \citealt{Mat08,CLPT,Whi14} or \citealt{VCW16} for the higher-order terms).  The large scale, Eulerian bias is simply related to our Lagrangian bias parameter ($b_1$) through $b=1+b_1$.

At this point we have two routes forward, and we shall discuss them in the next two subsections.
Section \ref{sec:za} presents a numerical calculation, whereas in Section \ref{sec:ZAx2} we will expand the Zeldovich calculation to $\mathcal{O}(x^2)$ to make contact with the linear theory solution.

\subsection{Numerical evaluation}
\label{sec:za}

The first approach is to simply evaluate the integral in Eq.~(\ref{eqn:ZelXi}) numerically.  We start by considering the bisector definition of $\hat{d}$.  Placing $\hat{d}$ along the $\hat{z}$-axis and orienting $\hat{s}_1$ and $\hat{s}_2$ in the $x-z$ plane the 3D integral is well behaved and converges rapidly for any triangle configuration.  A further numerical integral over $\mu$ at fixed $s$ and $d$ then gives $\xi_\ell(s,d)$.  We show this in Fig.~\ref{fig:multipoles_ZA} for $\ell=0$, 2 and 4 for several values of $d$.  In order to isolate the wide-angle behavior from the evolving dynamics we have assumed (unphysically) a fixed $\Lambda$CDM model at $z=0.25$ for each of these situations and set $b_1=0$.  In reality increasing depth ($d$) would also change the mean redshift, the growth and the degree of non-linearity.  These effects can all be accounted for, on large scales, by the ZA.
Figs.~\ref{fig:multipoles_b1} and \ref{fig:multipoles_b2} show the ratio of the correction terms to the ``plane-parallel'' limit (Eq.~\ref{eq:xiK}), and compares the corrections in linear theory and the ZA.
For the monopole and the quadropole our results are close to linear theory (presented in \citealt{CasWhi18}), but for $\ell=4$ we find a bigger difference on BAO scales with respect to the linear case.

A comparison of Figs.~\ref{fig:multipoles_b1} and \ref{fig:multipoles_b2} shows the influence of the bias terms.  Since we are treating the bias perturbatively, following \citet{Mat08}, we expect this calculation to be most accurate at large scales.
In linear theory and in plane-parallel approximation, biasing boils down to replacing $f\rightarrow f/b$ and $\xi_\ell^{(0)}\rightarrow b^2\xi_\ell^{(0)}$ , which implies that bias suppresses redshift-space effects for $\ell=0$ and $2$ but not for $\ell=4$.
\fig{fig:multipoles_b2} shows the ratio between the multipoles of the correlation function in linear theory and the ZA for a tracer with $b=1+b_1=2$ to the ones in the plane parallel limit. As expected, for $\ell=0$ the wide angle contributions are suppressed compared to the dark matter only case, but they are enhanced for $\ell=4$ where they can reach the $10\%$ level near the acoustic scale ($s\sim 110\,h^{-1}$Mpc).

It is clear from Eq.~(\ref{eqn:ZelXi}) that the wide-angle effects enter the bias terms differently than they do the matter terms.  To further investigate the effect of biasing on wide angle redshift space distortions in the ZA, Fig.~\ref{fig:multipoles_bZA} compares the full result for halos and matter for $\ell=0$, 2, 4. As discussed above, in the plane-parallel approximation the hexadecapole of halos and matter is the same in linear theory and we therefore expect it to be more sensitive to wide angle effects. This is precisely what Fig.~\ref{fig:multipoles_bZA} shows.

\begin{figure*}
\resizebox{0.85\columnwidth}{!}{\includegraphics{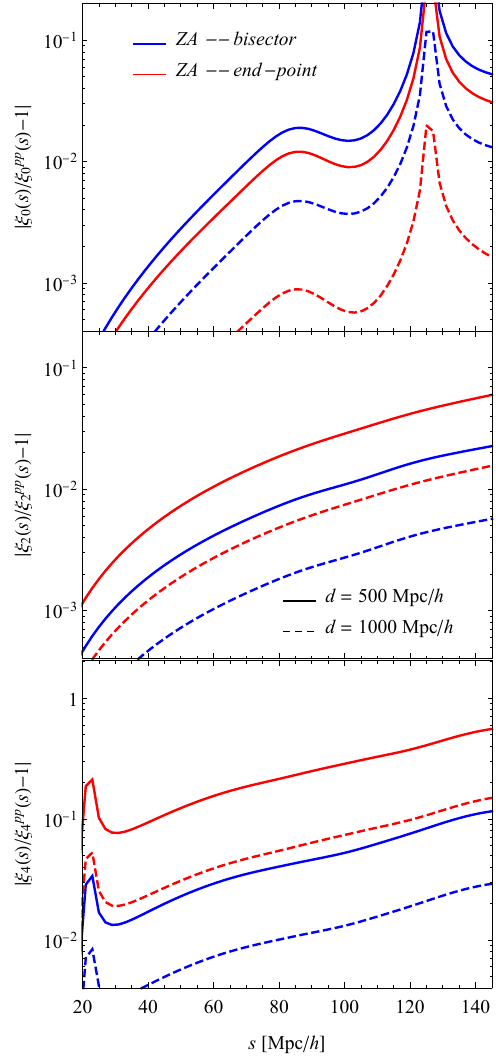}}\hfil
\resizebox{0.85\columnwidth}{!}{\includegraphics{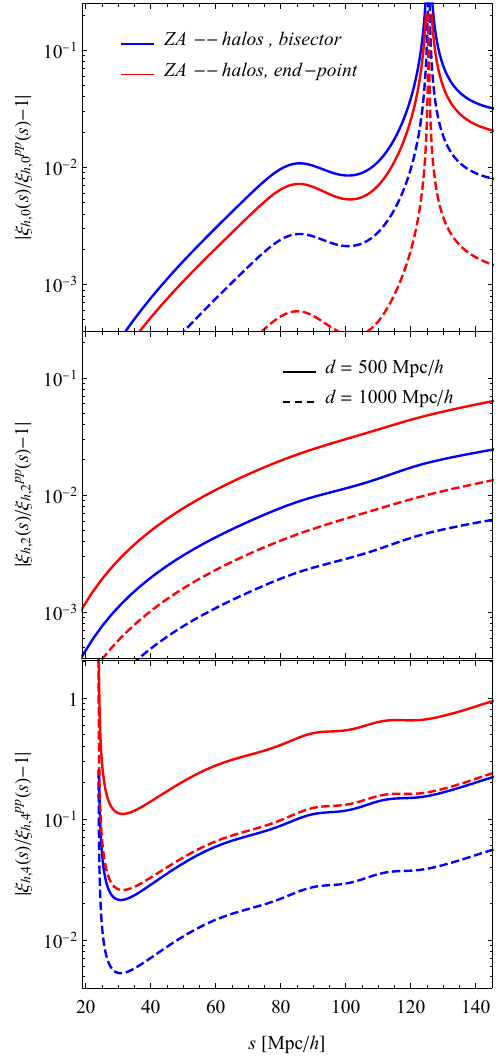}}
\caption{Same as Fig.~\ref{fig:multipoles_b1} but now using the end-point approximation $\hat{d}\approx\hat{s}_1$ instead of the bisector.  The left panels (for $\ell=0$, 2 and 4 from top to bottom) show the matter predictions ($b=1$) while the right panels show the halos with $b=2$.}
\label{fig:multipoles_endpoint}
\end{figure*}

\begin{figure}
\begin{center}
\resizebox{0.95\columnwidth}{!}{\includegraphics{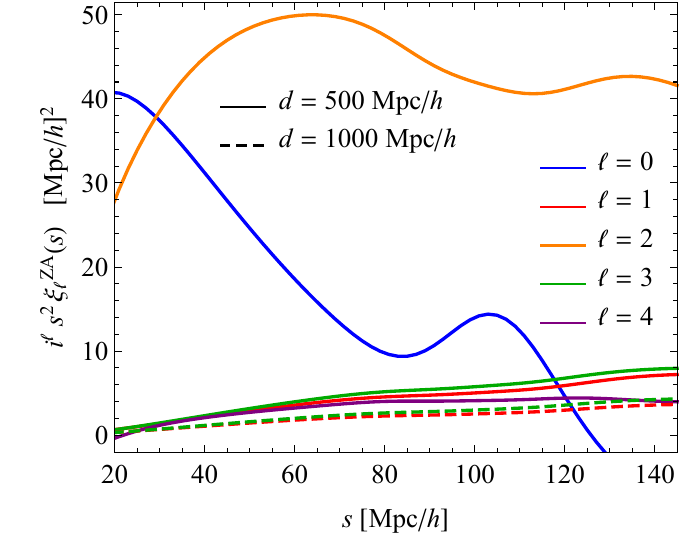}}
\end{center}
\caption{The dipole and octupole of the matter correlation function in the case of the end-point approximation for $s_1=500\,h^{-1}$Mpc and $1\,h^{-1}$Gpc, compared to the the even multipoles in the flat-sky limit.}
\label{fig:multipoles_odd}
\end{figure}

Next we turn to the end-point approximation.  Here we hold $s_1$ and $s$ fixed and integrate over $\mu_1$ to define the multipoles $\xi_\ell(s,s_1)$.  This is no more difficult than the bisector case, numerically, since we can compute $\xi(\vec{s}_1,\vec{s}_2)$ with ease for any triangle configuration.  However since this breaks the symmetry inherent in the bisector definition the corrections are larger.  This is shown in Fig.~\ref{fig:multipoles_endpoint} where we see the corrections becoming tens of percent at large scales.  In the end-point approximation we also generate odd multipoles, comparable in size with $\ell=0,4$ multipoles above $s = 120 \Mpc$ (Fig.~\ref{fig:multipoles_odd}).
The odd terms result from our choice of coordinate system, and therefore are not real physical effects, and should not be confused with relativistic dipoles present in the cross-correlation between two different tracers \citep{Bonvin14,Vid16,Lepori18}. Nevertheless they should be taken into account in the search for GR effects.
The trends seen in Figs.~\ref{fig:multipoles_endpoint} and \ref{fig:multipoles_odd} can be explained by expanding our expressions in powers of $x$, as we shall do in the next section.

\subsection{Small angle expansion}
\label{sec:ZAx2}

The second route is to expand the Zeldovich expression in \eq{eqn:ZelXi} in powers of $x$ and look at the correction terms analytically in order to gauge their structure. This is useful since it allows us to better understand the differences with linear theory and in which limits the latter is recovered. Again let us begin with the expressions when $\hat{d}$ is taken to be the angle bisector.  To this end, let us write
\begin{align}
  R_{2,ij} &= \left(1+f\hat{s}_2\hat{s}_2\right)_{ij} \\
      &= \left(1+\frac{f}{s_2^2}\left[\vec{d}-t\vec{s}\right]
                                \left[\vec{d}-t\vec{s}\right]\right)_{ij} \\
      &= \left(1+f\hat{d}\hat{d}\right)_{ij} + fx\left[\mu\hat{d}_i\hat{d}_j - \frac{\hat{d}_i\hat{s}_j+\hat{d}_j\hat{s}_i}{2}\right] \nonumber \\
      &+ \frac{fx^2}{4}\left[\hat{s}_i\hat{s}_j-\mu\hat{s}_i\hat{d}_j-\mu\hat{s}_j\hat{d}_i+(2\mu^2-1)\hat{d}_i\hat{d}_j \right] \nonumber \\
      &+ \mathcal{O}(x^3) \\
      &\equiv \sum_n x^n R_{2,ij}^{(n)}
\end{align}
where we have used $s_2^2=d^2-2tds\mu+t^2s^2=d^2(1-2tx\mu+t^2x^2)$ and the expressions in Section \ref{sec:background}.  Similarly
\begin{align}
  R_{1,ij} &= \left(1+f\hat{s}_1\hat{s}_1\right)_{ij} \\
      &= \left(1+f\hat{d}\hat{d}\right)_{ij} - fx\left[\mu\hat{d}_i\hat{d}_j - \frac{\hat{d}_i\hat{s}_j+\hat{d}_j\hat{s}_i}{2}\right] \nonumber \\
      &+ \frac{fx^2}{4}\left[\hat{s}_i\hat{s}_j-\mu\hat{s}_i\hat{d}_j-\mu\hat{s}_j\hat{d}_i+(2\mu^2-1)\hat{d}_i\hat{d}_j \right] \nonumber \\
      &+ \mathcal{O}(x^3)
\end{align}
Note that $R_{ij}^{(1)}\equiv R_{1,ij}^{(1)}=-R_{2,ij}^{(1)}$ while $R_{1,ij}^{(2)}=R_{2,ij}^{(2)}$.
Since $A_{ij}$ is symmetric, we see that the $\mathcal{O}(x)$ terms in $A_s$ vanish\footnote{A similar cancellation of the $\mathcal{O}(x)$ terms occurs in $U_{s,i}$.} (as we expect; \citealt{Sza98,Rei16,CasWhi18}) and we are left with
\begin{align}
 A_{s,ij} &= A_{\infty,ij}  + 2 x^2\Sigma^2[+ R_{in}^{(1)}R_{nj}^{(1)} +  2R_{in}^{(0)}R_{nj}^{(2)} ] \nonumber \\
  &+  2 x^2[-R_{im}^{(1)} \xi_{mn} R_{nj}^{(1)} + 2R_{im}^{(0)} \xi_{mn} R_{nj}^{(2)} ]+ \cdots
\end{align}
where we have used $R_{ij}^{(n)}=R_{ji}^{(n)}$ for $n=0$ and $2$.  Note the $x^2$ correction only has support for $\mu^0$, $\mu^1$ and $\mu^2$, but this couples to the existing $\mu$-dependence of the $x^0$ terms to create a richer structure.
Again it is important to note that in the $d^3q$ integral defining $1+\xi$ the values of $f$, $x$, $\mu$, $\hat{d}$ and $\hat{s}_i$ are constant.

To continue with the approximations, in order to gain some analytic intuition for the wide angle effects, let us follow \citet{Mat08} and in the decomposition $A_{ij}=2[\Sigma^2\delta_{ij}-\xi_{ij}(q)]$, treat $\xi_{ij}$ as small.  This is not optimal on smaller scales \citep{CLPT}, but will suffice to gain intuition on very large scales where the wide angle effects are largest.  In the same spirit, we have also dropped the term in Eq.~(\ref{eqn:ZelXi}) going as $U_iU_j$, which is $\mathcal{O}(P_L^2)$.

Starting from our expression for the correlation function, Eq.~(\ref{eqn:ZelXi}), we can pull the $q$-independent piece of $\exp[-(1/2)k_iA_{s,ij}k_j]$ out of the integral.  The plane-parallel term is
\begin{align}
  \mathcal{D}^{(0)} &= \exp\left[-\Sigma^2 k_i R_{im}^{(0)}R_{mj}^{(0)}k_j\right]  \\
  &= \exp\left[-k^2\Sigma^2(1+f[f+2]\mu_k^2) \right]
\end{align}
where $\mu_k=\hat{k}\cdot\hat{d}$.  This agrees with the form derived in \citet{Mat08} and can be rewritten as
\begin{equation}
  \mathcal{D}^{(0)}=\exp\left[-k_\parallel^2\Sigma_\parallel^2-k_\perp^2\Sigma_\perp^2\right]
\end{equation}
if $k_\perp^2=k^2(1-\mu_k^2)$, $k_\parallel=k\mu_k$, $\Sigma_\perp=\Sigma$ and $\Sigma_\parallel=(1+f)\Sigma$.  This term is responsible for the broadening of the BAO peak in the correlation function.  The $\mathcal{O}(x^2)$ correction is
\begin{align}
& 1+x^2\mathcal{D}^{(2)}
= \exp\left[-x^2\Sigma^2 k_i\left(R_{in}^{(1)}R_{nj}^{(1)}+2R_{in}^{(0)}R_{nj}^{(2)} \right) k_j + \cdots\right] \\ & \simeq 1+x^2  k^2 \Sigma^2 \frac{1}{4} f (f+2) \left(\left(2 \mu ^2-1\right) \mu _k^2-2 \mu  \nu 
   \mu _k+\nu ^2\right) 
\end{align}
where we have defined $\nu=\hat{k}\cdot\hat{s}$.
Note that this suggests that the broadening is opening angle dependent but that the correction is generally small.

Continue by expanding $\xi_{ij}(q)$ out of the exponential and doing the $d^3q$ integral.  The lowest order term regains the usual expression \citep{Kai87,Mat08}
\begin{align}
  \hat{k}_i R_{ij}^{(0)}\hat{k}_j\ \hat{k}_m R_{mn}^{(0)}\hat{k}_n\ P(k)
  &= \mathcal{K}^{(0)}\ P(k) \\
  &= (1+f\mu_k^2)^2\ P(k)
\label{eqn:KaiserLimit}
\end{align}
while the first correction for the matter is $x^2P$ times
\begin{align}
\label{eq:K2}
\mathcal{K}^{(2)} &\equiv -\left(\hat{k}_i R_{ij}^{(1)}\hat{k}_j\right)^2
     + 2\hat{k}_i R_{ij}^{(0)}\hat{k}_j\ \hat{k}_m R_{mn}^{(2)}\hat{k}_n \\
  &= -f^2\mu_k^2\left(\mu_k \mu -\nu\right)^2 \nonumber \\ 
  &+ (1+f\mu_k^2) \times \frac{f}{2}\left[\nu^2-2\mu\mu_k\nu+(2\mu^2-1)\mu_k^2\right] \notag \\
  &= \frac{f}{2}\left\{ \frac{1}{3}\left(1-[1+f]\mu_k^2-3f\mu_k^4\right) \right. \nonumber \\
  &- \mathcal{L}_1(\mu)\mathcal{L}_1(\nu)\,2\mu_k\left(1-f\mu_k^2\right) \nonumber \\
  &+ \left. \mathcal{L}_2(\mu)\frac{4}{3}\mu_k^2 + \mathcal{L}_2(\nu)\frac{2}{3}[1-f\mu_k^2] \vphantom{\int}\right\}
\end{align}
The bias terms are enumerated in Appendix \ref{sec:bias}. It is  however important to note that the leading wide angle correction to the bias terms, \eq{eq:b1WA}, is different than the one described above as it does not receive a contribution $\propto \left(k_iR_{ij}^{(1)}k_j\right)^2$.
Our expression, for the matter, is thus
\begin{align}
  \xi^{(s)}(s,d,\mu)
  &= \int\frac{\mathrm{d}^3k}{(2\pi)^3} e^{i\vec{k}\cdot\vec{s}} \mathcal{D}^{(0)}(k,\mu_k) \notag \\ 
  &\times P(k)\left[\mathcal{K}^{(0)} + x^2 \mathcal{K}^{(2)}\right]\left(1+x^2 \mathcal{D}^{(2)}\right)
\label{eqn:ZelMatXi}
\end{align}
The $\mathcal{O}(x^0)$ term is simply the normal Kaiser expression with $P\to\mathcal{D}^{(0)}P$.  The $\mathcal{O}(x^2)$ terms give the leading wide angle correction.

Before we study the $\mathcal{O}(x^2)$ terms let us quickly review the calculation to lowest order:
\begin{align}
  \xi_\ell
  &\ni (2\ell+1)\int\frac{\mathrm{d}^3k}{(2\pi)^3}\frac{\mathrm{d}^2\hat{s}}{4\pi}
  \ e^{i\vec{k}\cdot\vec{s}}
  \ \mathcal{L}_\ell(\hat{s}\cdot\hat{d}) \nonumber \\
  &\times \mathcal{D}^{(0)}(k,\mu_k)\left(1+f\mu_k^2\right)^2P(k) \\
  &= (2\ell+1)i^\ell\int\frac{k^2\mathrm{d}k}{2\pi^2}\ P(k)j_\ell(ks)\nonumber \\
  &\times \int\frac{\mathrm{d}\Omega_k}{4\pi}\mathcal{L}_\ell(\mu_k)\mathcal{D}^{(0)}(k,\mu_k)\left(1+f\mu_k^2\right)^2
\label{eqn:xiell_x0}
\end{align}
where $\mu_k=\hat{k}\cdot\hat{d}$.  In the absence of the damping term, the $d\Omega_k$ integral gives the usual multipoles $\ell=0$, 2 and 4 and one recovers Eq.~(\ref{eq:xiK}).  The anisotropic damping also populates $\ell>4$.  While the results of the $d\Omega_k$ integral can be written in closed form, they are not illuminating and so will be omitted \citep[see e.g.][for further discussion]{PD94}.

Now we consider the $\mathcal{O}(x^2)$ terms.  For these terms $\xi_\ell$ has $\hat{s}$ dependence not just through $\exp[i\vec{k}\cdot\vec{s}]$ and $\mathcal{L}_\ell(\hat{s}\cdot\hat{d})$ but also through the $\mu$ and $\nu$ terms in Eq.~(\ref{eqn:ZelMatXi}).  However the terms are at most quadratic in these variables.  The $\mathcal{O}(x^2)$ contribution to $\xi_\ell$ is
\begin{align}
  \xi_\ell
  &\ni (2\ell+1)x^2 \int\frac{\mathrm{d}^3k}{(2\pi)^3}\frac{\mathrm{d}\hat{s}}{4\pi} \ e^{i\vec{k}\cdot\vec{s}}\mathcal{L}_\ell(\hat{s}\cdot\hat{d}) \nonumber \\
  &\times \mathcal{D}^{(0)}(k,\mu_k)\,P(k) \left\{\mathcal{K}^{(0)}\mathcal{D}^{(2)} + \mathcal{K}^{(2)} \right\}
\end{align}
Upon performing the integral over $\mathrm{d}^2\hat{s}$ using the expressions in Appendix \ref{app:x2_terms}, the contribution is of the form
\begin{align}
 \xi_\ell &\ni (2\ell+1)x^2\int\frac{k^2\mathrm{d}k}{2\pi^2}\ P(k) \sum_L i^L j_L(ks) \nonumber \\
 &\times \int\frac{\mathrm{d}\Omega_k}{4\pi}\ \mathcal{C}_L^\ell(\mu_k)\mathcal{D}^{(0)}(k,\mu_k)
\label{eqn:CLl}
\end{align}
where the  $\mathcal{C}_L^\ell$ are polynomials in $\mu_k$.
We give the general expressions for $\mathcal{C}_L^\ell$ in Appendix \ref{app:x2_terms}.  As an example
\begin{align}
   & C_0^0 =\frac{1}{12} f (f+2) k^2 \Sigma ^2 \left(1-3 \mu _k^2\right) \left(f \mu
   _k^2+1\right)^2 \\ & -\frac{1}{6} f \left(f \mu _k^4+(f+3) \mu _k^2-1\right) 
\end{align}
The polynomials for high $\ell$ are long and we shall not reproduce them here. Similar to the plane parallel limit in \eq{eqn:xiell_x0} the angular integral in \eq{eqn:CLl} can be evaluated analytically, but the resulting expression is not very illuminating.
The important point to note is that in general $\xi_\ell$ now contains contributions $j_L(ks)$ for $L\ne\ell$, as was the case for linear theory \citep[see the discussion in][]{CasWhi18}.  In fact, in the limit $\mathcal{D}^{(2)}\to 0$ and $\mathcal{D}^{(0)}\to 1$ we regain the earlier, linear theory, results.  This can be seen as an alternative route to those results, first derived by \citet{Sza98}.

It is straightforward to convert these\footnote{The conversions also hold for the linear theory results presented in \citet{CasWhi18}.} bisector-based results to the case where $\hat{d}$ is approximated by $\hat{s}_1$, i.e.~the end-point approximation.  The lowest order terms (Eq.~\ref{eqn:xiell_x0}) are unchanged and for the terms in Eq.~(\ref{eqn:CLl}), which are already $\mathcal{O}(x^2)$, we can simply replace $d$ with $s_1$ (i.e.~$x\to x_1$).  The change $\mu\to\mu_1$ mixes multipoles since
\begin{equation}
\label{eq:Lmu1}
  \mathcal{L}_\ell(\mu) = \mathcal{L}_\ell(\mu_1) + \frac{x_1}{2}\sqrt{1-\mu_1^2}\mathcal{L}^1_\ell(\mu_1)-\frac{x_1^2}{8}(1-\mu_1^2)\mathcal{L}^2_\ell(\mu_1)
\end{equation}
through $\mathcal{O}(x_1^2)$.
This populates the odd $\ell$.  Thus to the wide-angle bisector terms we must add the terms in Table \ref{tab:endpoint_terms} times $\xi_L^{pp}$.  Denoting the end-point expansion with a tilde, as for $\tilde{P}_\ell^{(n)}(k)$, we have for example $\tilde{\xi}_0(s)=\xi_0(s) + (x_1^2/5)\xi_2^{pp}(s)$ with $\xi_0(s)$ the bisector expression including the $\mathcal{O}(x^2)$ terms and $\xi_2^{pp}$ the plane-parallel limit.  Since $\xi_2^{pp} <0$, the formulae above explain why in \fig{fig:multipoles_endpoint} the end-point monopole is less affected by wide angle effects than the bisector-define monopole.
Table \ref{tab:endpoint_terms} explicitly shows that odd multipoles do not carry any other extra-information, as they are proportional to the multipoles in the plane-parallel limit, \eg $\tilde{\xi}_1(s,s_1) = -3/5 x_1 \xi^{pp}_2(s)$.

\begin{table}
\begin{center}
\begin{tabular}{ccc}
$\ell$ & $L=2$         & $L=4$ \\ \hline
 0     & $x_1^2/5$   & 0 \\
 1     & $-3x_1/5$   & 0 \\
 2     & $-2x_1^2/7$ & $5x_1^2/7$ \\
 3     & $3x_1/5$    & $-10x_1/9$ \\
 4     & $3x_1^2/35$ & $-90x_1^2/77$
\end{tabular}
\end{center}
\caption{The coefficients of the additional contributions to $\xi_\ell$ which are generated by the use of the end-point approximation ($\hat{d}\simeq\hat{s}_1$) as described in the text.}
\label{tab:endpoint_terms}
\end{table}

\subsection{Power spectrum}

We can express the expectation value of the Yamamoto estimator for the power spectrum as an integral over $\xi(s,d,\mu)$ and use Eq.~(\ref{eqn:ZelMatXi}) to study the impact of the wide angle terms, on large scales and to $\mathcal{O}(x^2)$.  Using $\hat{d}$ as the line of sight, the Yamamoto estimator is simply the Hankel transform of our correlation function multipoles:
\begin{align}
\left\langle P_L^Y(k)\right\rangle
&=(2L+1) \int\frac{\mathrm{d}\Omega_\mathbf{k}}{4\pi}\, \frac{\mathrm{d}^3d}{V} \,\mathrm{d}^3s\ e^{-i\mathbf{k}\cdot \mathbf{s}} \mathcal{L}_L(\hat{k}\cdot\hat{d})\xi(s,d,\mu) \notag \\
&=(2L+1) \int\frac{\mathrm{d}\Omega_\mathbf{k}}{4\pi}\, \frac{\mathrm{d}^3d}{V}\,\mathrm{d}^3s \ e^{-i\mathbf{k}\cdot \mathbf{s}}\mathcal{L}_L(\hat{k}\cdot\hat{d}) \notag \\
& \times\sum_{\ell} \xi_\ell(s,d)\mathcal{L}_\ell(\hat{s}\cdot\hat{d}) \\
&= (4\pi)(-i)^L\int\frac{\mathrm{d}^3d}{V}\,s^2\mathrm{d}s\ j_L(ks) \xi_L(s,d)
\end{align}
as in Eq.~(\ref{eq:Pkell}).  The lowest order terms simplify upon using the completeness relation
\begin{equation}
  \int s^2\,\mathrm{d}s\ j_\ell(ks)j_\ell(k's) = \frac{\pi}{2kk'}\ \delta^{(D)}(k-k') \quad .
\end{equation}
The $\mathrm{d}s$ integral times the Bessel function ``undoes'' the $\mathrm{d}k$ integral times the Bessel function in Eq.~(\ref{eqn:xiell_x0}) and
\begin{align}
  \left\langle P_L^{Y,(0)}(k)\right\rangle &= (2L+1)P(k)\nonumber \\ 
  &\times\int\frac{\mathrm{d}\Omega_k}{4\pi} \mathcal{D}^{(0)}(k,\mu_k)\mathcal{L}_L(\mu_k)\left(1+f\mu_k^2\right)^2
\end{align}
which recovers the expression in \citet{Mat08}.
Note the well-known exponential damping of the Zeldovich power spectrum.
Additional high-$k$ power is generated by contributions which are isolated to small $r$ in configuration space.  Various models for this missing power have been proposed, either heuristic \citep{ESW07,SSEW08}, based on the halo model \citep{MohSel14,SelVla15} or on effective field theory \citep{Por14,VWA15,VCW16}.

At second order plugging \eq{eqn:CLl} into \eq{eq:Pkell} does not further simplify since the integral of $j$'s of different orders does not vanish.  We are thus left with
\begin{align}
  \left\langle P_L^Y(k)\right\rangle &\ni (2L+1)x^2\int\frac{\mathrm{d}^3d}{V}\frac{\mathrm{d}^3k'}{(2\pi)^3}\ P(k')\mathcal{D}^{(0)}(k',\mu_k') \nonumber \\
  &\times \sum_J  C_J^L(\mu_k')\ (4\pi)\int s^2\mathrm{d}s\ j_L(ks)j_J(k's) \quad .
\end{align}
The last integral, over $\mathrm{d}s$, can be expressed analytically using hypergeometric functions, but the final expression does not provide any
further insights. 

The fast FFT estimator in \eq{eq:PkFFT} can be expressed in a similar form using either the bisector or the end-point as the LOS. The former has been presented in Eq.~32 of \citep{CasWhi18}, while the latter can be obtained from Eqs. 30-31 of \citep{CasWhi18} using the mapping in \eq{eq:Lmu1} and Table \ref{tab:endpoint_terms}.

\section{Conclusions}
\label{sec:conc}

The physics of electromagnetic emission from moving objects, which imprints a contribution from the line-of-sight peculiar velocity onto the observed redshift of extragalactic objects, breaks the translational invariance of our theories down to a rotational symmetry.  The induced effects, which become important in 2-point clustering statistics when the opening angle between the two points becomes appreciable, go under the name of ``wide angle effects''.  Since these effects are largest on large scales, most earlier papers have assumed Eulerian, linear perturbation theory in their analyses.  In this paper we have shown that wide angle effects can be easily handled within the context of Lagrangian perturbation theory, allowing an efficient resummation of the linear displacements which is particularly important for modeling BAO.

Beyond the plane-parallel approximation the two point function is most easily expressed in terms of the correlation function or the multi-frequency angular power spectrum (MAPS).  We developed the relationship between these probes and showed how the MAPS can be computed beyond linear theory.  We investigated the relationship of these statistics to the multipole moments of the power spectrum computed with the Yamamoto estimator, using either the bisector or end-point conventions for the line of sight direction.

We have compared our calculation, numerically and analytically, to the earlier linear theory calculations.  Except near the BAO peak, where linear theory does a poor job, the size of the corrections for the $\ell=0$ and $2$ multipole moments of the correlation function are very similar in the ZA and in linear theory.  For $\ell=4$ the corrections predicted in the ZA are larger than the linear theory predictions.  The corrections are significantly larger if the end-point convention is used to define the line of sight than if the bisector approximation is made.

We note that it is relatively straightforward, if tedious, to extend our analysis to higher order in (Lagrangian) perturbation theory.  This would allow a comparison of the size of the wide angle terms to those from second order dynamics.  What is much more difficult is an extension of this work to schemes such as the streaming model \citep[e.g.][and references therein]{VCW16}, in which a fixed line of sight is critical to the simplification of the final expressions.  However, on small scales the wide-angle terms are small while on large scales the corrections to the dynamics are small.  This suggests a perturbative approach where the wide-angle corrections are computed at low order (as we have done here) and used to correct the more sophisticated model, that is computed in the plane-parallel approximation.

\vspace{0.2in}
M.W.~is supported by the U.S.~Department of Energy and by NSF grant number 1713791.
This work made extensive use of the NASA Astrophysics Data System and of the {\tt astro-ph} preprint archive at {\tt arXiv.org}. 

\appendix

\section{Flat sky approximation to the MAPS}
\label{app:flatsky_MAPS}

In the main text we discussed the relationship of the MAPS, $C_\ell(s_1,s_2)$, to the correlation function and power spectrum.  If we make the small-angle, or flat-sky, approximation and define $s_\parallel=s\mu$ and $s_\perp=s\sqrt{1-\mu^2}$ then
\begin{align}
  C_\ell(s_1,s_2) &= 2\pi\,\int \mathrm{d}(\cos\theta)\ \xi(s,d,\mu)\mathcal{L}_\ell(\cos\theta) \\
  &\simeq 2\pi\int \widetilde{\omega}\mathrm{d}\widetilde{\omega}\ \xi(s_\perp,s_\parallel,d)J_0(\ell\widetilde{\omega})
\end{align}
where $\widetilde{\omega}=2\sin(\theta/2)\simeq\theta$.  Changing arguments to $s_1-s_2\simeq s_\parallel + \mathcal{O}(x^3)$ and $\frac{1}{2}(s_1+s_2)\simeq d+\mathcal{O}(x^2)$ and writing $\ell=k_\perp d$ so  $\ell\widetilde{\omega} \simeq k_\perp s_\perp$ we find
\begin{align}
  C_\ell(s_\parallel,d) &\simeq  \int\frac{\mathrm{d}^2s_\perp}{d^2}\ \xi(s_\parallel,s_\perp,d)e^{i\vec{k}_\perp\cdot\vec{s}_\perp}
\end{align}
where we have used the Rayleigh expansion of the plane-wave in cylindrical coordinates and the azimuthal symmetry of the integral.
Thus the MAPS, in the flat-sky limit, is the 2D Fourier transform of the correlation function.  A further Fourier transform (in $s_\parallel$) returns $P(\vec{k})$.  Alternatively the MAPS is the line-of-sight Fourier transform of $P(k_\parallel,k_\perp)$:
\begin{equation}
  C_{\ell}(s_\parallel,d) \simeq \int_0^\infty \frac{dk_\parallel}{\pi\,d^2}
  \ P(k_\perp=\ell/d,k_\parallel)\cos\left(k_\parallel s_\parallel\right) \quad .
\end{equation}
For an alternative derivation, at the level of the fields, see the appendices of \citet{WCDH99}, \citet{Dat07} or \citet{WhiPad17}.

\section{Bias terms}
\label{sec:bias}

The low-$k$ expansion of the bias terms in Eq.~(\ref{eqn:ZelXi}) follows very similar steps to the one for the matter terms presented in the main text.  The $b_1^2$ term does not carry any extra redshift space dependence and therefore is identical to the expansion of $\xi_{ij}$ in Eq.~(\ref{eqn:KaiserLimit}).  The $b_1$ piece is also straightforward. 
Expanding $U_{s,i}$ to quadratic order in $x$ we get
\begin{align}
U_{s,i}(q) = R^{(0)}_{ij} U_j(q)+R^{(2)}_{ij} U_j(q) \equiv
U_{s,i}^{(0)}(q)+U_{s,i}^{(2)}(q)
\end{align}
which we can then plug back into Eq.~(\ref{eqn:ZelXi}).  At lowest order,
\begin{align}
  \xi_{b_1}^{(0)}(s,d,\mu) &= 2b_1\int\frac{\mathrm{d}^3k}{(2\pi)^3} e^{i\vec{q}\cdot\vec{s}}\mathcal{D}^{(0)}\int \mathrm{d}^3q \int\frac{\mathrm{d}^3p}{(2\pi)^3}e^{i\vec{q}\cdot(\vec{p}-\vec{k})} \notag \\
  &\times k_i \left(\delta_{ij} + f \hat{d}_i \hat{d}_j\right) i\frac{p_j}{p^2}P_L(p) \notag \\
  &= 2b_1\int\frac{\mathrm{d}^3k}{(2\pi)^3} \mathcal{D}^{(0)}e^{i\vec{q}\cdot\vec{s}}(1+f \mu_k^2)P_L(k)
\end{align}
and we recover the familiar Kaiser result
\begin{equation}
  P_{L,s}(k,\mu_k) = \left([1+b_1]+f \mu_k^2\right)^2P_L(k)
\end{equation}
upon recalling the large scale (Eulerian) bias is $b=1+b_1$. 
At second order
\begin{align}
\label{eq:b1WA}
  \xi_{b_1}^{(2)}(s,d,\mu) &= 2b_1x^2\int\frac{\mathrm{d}^3k}{(2\pi)^3} e^{i\vec{q}\cdot\vec{s}}\int \mathrm{d}^3q \int\frac{\mathrm{d}^3p}{(2\pi)^3}e^{i\vec{q}\cdot(\vec{p}-\vec{k})} \notag \\
  &\times \mathcal{D}^{(0)} k_i \left(R^{(2)}_{ij}+\mathcal{D}^{(2)}R^{(0)}_{ij}\right) i\frac{p_j}{p^2}P_L(p) \notag \\
 &= 2b_1x^2\int\frac{\mathrm{d}^3k}{(2\pi)^3} e^{i\vec{k}\cdot\vec{s}}P_L(k)\mathcal{D}^{(0)} \nonumber \\
 &\times \left[\frac{f}{4}\left(\nu^2-2\mu\mu_k\nu+\left(2\mu^2-1\right)\mu_k^2\right) \right.    \notag \\
  &+  \left. \mathcal{D}^{(2)}\left(1+f \mu_k^2\right)\right]
\end{align}
It is worth pointing out that leading order wide angle contribution calculated above differs from the dark matter one in \eq{eq:K2} even neglecting the expansion of the damping term. This is a new feature of the ZA and it explains why the ZA is more different from linear theory for halos than it is for dark matter.
\section{The wide angle terms}
\label{app:x2_terms}

As described in the main text, for the $\mathcal{O}(x^2)$ terms, $\xi_\ell$ has $\hat{s}$ dependence not just through $\exp[i\vec{k}\cdot\vec{s}]$ and $\mathcal{L}_\ell(\hat{s}\cdot\hat{d})$ but also through the $\mu$ and $\nu$ terms in Eq.~(\ref{eqn:ZelMatXi}).  However these are at most quadratic in these variables.  We rewrite each of the terms using e.g.~$\nu^2=(1/3)+(2/3)\mathcal{L}_2(\nu)$ and $\mu\nu=\mathcal{L}_1(\mu)\mathcal{L}_1(\nu)$.
 
The $\mathcal{O}(x^2)$ terms which are independent of $\mu$ and $\nu$ go through as for the $\mathcal{O}(x^0)$ terms, giving a Hankel transform of order $\ell$:
\begin{equation}
 \int\frac{\mathrm{d}\hat{s}}{4\pi}\ e^{i\vec{k}\cdot\vec{s}}\mathcal{L}_\ell(\hat{s}\cdot\hat{d})
    = i^\ell j_\ell(ks) \mathcal{L}_\ell(\hat{k}\cdot\hat{d}) \quad .
\end{equation}
The other terms are proportional to $\mathcal{L}_2(\nu)$, $\mathcal{L}_2(\mu)$ and $\mathcal{L}_1(\nu)\mathcal{L}_1(\mu)$ and will additionally give Hankel transforms of different orders: $L\ne\ell$. Using the angular momentum addition theorem for Legendre polynomials
\begin{equation}
  \mathcal{L}_{\ell_1}(\mu)\mathcal{L}_{\ell_2}(\mu) = \sum_L \tj{\ell_1}{\ell_2}{L}{0}{0}{0}^2
  (2L+1)\mathcal{L}_L(\mu)
\end{equation}
the Rayleigh expansion of a plane wave
\begin{equation}
  e^{i\vec{k}\cdot\vec{s}} = \sum_\ell i^\ell(2\ell+1)j_\ell(ks)
  \mathcal{L}_\ell(\hat{k}\cdot\hat{s})
\end{equation}
and the addition theorem
\begin{equation}
  \mathcal{L}_\ell(\hat{n}_1\cdot\hat{n}_2) = \frac{4\pi}{2\ell+1}
  \sum_m Y_{\ell m}(\hat{n}_1)Y_{\ell m}^\star(\hat{n}_2)
\end{equation}
one can show
\begin{align}
  & \int\frac{\mathrm{d}\hat{s}}{4\pi}\ e^{i\vec{k}\cdot\vec{s}}\mathcal{L}_\ell(\hat{s}\cdot\hat{d})
    \mathcal{L}_2\left(\hat{s}\cdot\hat{k}\right) \notag \\
    &= \sum_L \tj{L}{2}{\ell}{0}{0}{0}^2(2L+1)i^L j_L(ks) \mathcal{L}_\ell(\hat{k}\cdot\hat{d})
\end{align}
For each $\ell$ only a finite number of terms with $L\ne\ell$ contribute and $L$ is even. 
We also have
\begin{align}
  & \int\frac{\mathrm{d}\hat{s}}{4\pi}\ e^{i\vec{k}\cdot\vec{s}}\mathcal{L}_\ell(\hat{s}\cdot\hat{d})
    \mathcal{L}_2\left(\hat{s}\cdot\hat{d}\right) \notag \\
    &= \sum_L \tj{L}{2}{\ell}{0}{0}{0}^2 (2L+1)i^L j_L(ks) \mathcal{L}_L(\hat{k}\cdot\hat{d})
\end{align}
which also contains only even $L$.  Finally
\begin{align}
  & \int\frac{\mathrm{d}\hat{s}}{4\pi}\ e^{i\vec{k}\cdot\vec{s}}\mathcal{L}_\ell(\hat{s}\cdot\hat{d})
  \mathcal{L}_1\left(\hat{s}\cdot\hat{k}\right)\mathcal{L}_1\left(\hat{s}\cdot\hat{d}\right) \notag \\
    &= \sum_{JL}\tj{\ell}{1}{J}{0}{0}{0}^2 \tj{L}{1}{J}{0}{0}{0}^2  (2L+1)(2J+1)i^L \notag \\
  &\times j_{L}(ks) \mathcal{L}_J(\hat{k}\cdot\hat{d})
\end{align}
The double sum is also finite and contains only even $L$'s.
The $\mathcal{O}(x^2)$ contribution to $\xi_\ell$ is then
\begin{align}
  &  (2\ell+1)x^2\int\frac{\mathrm{d}^3k}{(2\pi)^3}\frac{\mathrm{d}\hat{s}}{4\pi} e^{i\vec{k}\cdot\vec{s}} \mathcal{L}_\ell(\mu)\mathcal{D}^{(0)}\,P(k) \nonumber \\
  & \left\{ T_{00} + T_{11}\mathcal{L}_1(\mu)\mathcal{L}_1(\nu) + T_{20}\mathcal{L}_2(\mu) + T_{02}\mathcal{L}_2(\nu) \vphantom{\int}\right\} \\
  &= x^2\int\frac{\mathrm{d}^3k}{(2\pi)^3}\mathcal{D}^{(0)}(k,\mu_k)\,P(k)\sum_L i^Lj_L(ks) \nonumber \\
  &\times (2\ell+1) \left\{ T_{00}\delta_{L\ell}\mathcal{L}_\ell(\mu_k) \vphantom{\int} \right. \nonumber \\
  &+        T_{11}\sum_J (2L+1)(2J+1) \tj{L}{1}{J}{0}{0}{0}^2 \tj{\ell}{1}{J}{0}{0}{0}^2 \mathcal{L}_J(\mu_k) \nonumber \\
  &+        T_{20}(2L+1) \tj{\ell}{2}{L}{0}{0}{0}^2\mathcal{L}_L(\mu_k) \nonumber \\
  &+ \left. T_{02}(2L+1) \tj{L}{2}{\ell}{0}{0}{0}^2\mathcal{L}_\ell(\mu_k)\right\}
\end{align}
where
\begin{align}
& T_{00} =  \frac{1}{12} f \left((f+2) k^2 \Sigma ^2 \left(\mu _k^2-1\right) \left(f \mu
   _k^2+1\right)^2\right.\\
  & \left.-6 f \mu _k^4-2 (f+1) \mu _k^2+2\right) \\ 
        & T_{11} = \frac{1}{2} f \mu _k \left(-(f+2) k^2 \Sigma ^2 \left(f \mu
   _k^2+1\right){}^2+2 f \mu _k^2-2\right)\\
&  T_{20} = \frac{1}{3} f \left((f+2) k^2 \Sigma ^2 \left(f \mu _k^3+\mu _k\right){}^2+2
   \mu _k^2\right)\\
  &T_{02} = \frac{1}{6} \left(f (f+2) k^2 \Sigma ^2 \left(f \mu _k^2+1\right)^2-2 f
   \left(f \mu _k^2-1\right)\right)
\end{align}
The triangle condition on the $3j$ symbols ensures that only a finite number of terms contribute for any $\ell$, and it is straightforward to compute $\mathcal{C}_L^\ell$ of the main text from the above expressions.  The symmetry of the problem ensures that the dipole terms which one might naively think appear in the sum in fact cancel exactly.

\bibliography{main}

\end{document}